\documentstyle[amssymb,prl,aps,epsf]{revtex}
\begin{document}
\title{Quasiparticles as composite objects in the RVB superconductor}
\author{Yi Zhou$^a$, V. N. Muthukumar$^b,$ and Zheng-Yu Weng$^a$}
\address{$^a$ Center for Advanced Study, Tsinghua University, Beijing
100084, China\\
$^b$ Department of Physics, Princeton University, Princeton, NJ 08544} 
\maketitle
\begin{abstract}
We study the nature of the superconducting state, the origin of $d$-wave
pairing, and elementary excitations of a resonating valence bond (RVB)
superconductor. We show that the phase string formulation of the $t-J$ model
leads to confinement of bare spinon and holon excitations in the
superconducting state, though the vacuum is described by the RVB state.
Nodal quasiparticles are obtained as composite excitations of spinon and
holon excitations. The $d$-wave pairing symmetry is shown to arise from
short range antiferromagnetic correlations.
\end{abstract}



\section{Introduction}

This paper concerns the nature of the superconducting state and the
excitation spectrum obtained from the bosonic resonating valence bond
(b-RVB) theory of the $t-J$ model. The $t-J$ model, arguably the simplest
description of a doped Mott insulator \cite{gros_87}, is given by 
\begin{equation}
H=-t\sum_{\langle ij\rangle \sigma }c_{i\sigma }^{\dagger }(1-n_{i-\sigma
})c_{j\sigma }(1-n_{j-\sigma })+J\sum_{\langle ij\rangle }~\left[ \mathbf{S}%
_i\cdot \mathbf{S}_j-\frac{n_in_j}4\right] .  \label{htj}
\end{equation}
The b-RVB theory of the Hamiltonian (\ref{htj}) yields a consistent
description of the doped antiferromagnetic Mott insulator\cite{zy_99}. At
half filling, the theory reduces to the Schwinger boson representation of
the Heisenberg model \cite{arovas_88,sarker_89,yoshioka_89}. At small doping
concentrations, the theory describes a cluster spin glass phase, which
eventually makes way to a superconducting phase \cite{kou_02}. The theory
also exhibits a spontaneous vortex phase where the holons are (Bose)
condensed and the spinons are perceived as half vortices by the holon field 
\cite{zy_02}. Recently, we considered a Ginzburg-Landau formulation of this
theory and showed how $\frac{hc}{2e}$ quantization arises in the
superconducting state. A spinon, perceived as a half vortex by the holon
field, is trapped within a vortex core, thereby leading to a $2e$ flux
quantum \cite{vnm_02}. In this paper, we present a microscopic description
of the superconducting state at zero temperature. In particular, we show
explicitly that when the holons are condensed, it costs a logarithmically
divergent energy to create a bare spinon excitation. Therefore, a bare
spinon excitation is confined. We show that energy considerations only
permit neutral $S=1$ and quasiparticle excitations. We also demonstrate that
the short range antiferromagnetic (AF) correlations lead to $d$-wave pairing
symmetry. The results presented in this paper illustrate how a nodal
quasiparticle arises from an RVB vacuum, at low energies. This paper
complements the macroscopic description of the superconducting state,
presented in our earlier work \cite{vnm_02}.

The paper is organized as follows. For completeness, we devote the next
section to a brief summary of the b-RVB formulation of the $t-J$ model (\ref
{htj}). This leads to an effective Hamiltonian for the spin and charge
degrees of freedom which can be solved within a mean field approximation.
The basic results used repeatedly in this paper are summarized. In the
following section, we discuss the nature of the superconducting state, and
show how $d$ -wave pairing arises in this theory. The excitation spectrum is
studied in Section IV. We show that bare spinon (or holon) excitations are
confined in the superconducting state, and that only neutral $S=1$ and
composite excitations (that carry both spin and charge) are allowed in the
superconducting state. The latter would be the RVB analog of quasiparticles
in a BCS superconductor. In the final section, we summarize our results and
discuss them in the context of the complete phase diagram obtained within
the b-RVB theory.

\section{Bosonic formulation of the $t-J$ model}

The b-RVB formulation can be regarded as a bosonization scheme for the $t-J$
model (\ref{htj}). The electron operator, $c_{i\sigma}$ is written as 
\begin{equation}
c_{i\sigma }=h_i^{\dagger }b_{i\sigma }e^{i\hat{\Theta}_{i\sigma }}
\label{mutual}
\end{equation}
where the holon $h_{i\sigma }^{\dagger }$ and spinon $b_{i\sigma }$
operators represent bosonic fields. The holon and spinon fields satisfy the
constraint, $h_i^{\dagger }h_i+\sum_\sigma b_{i\sigma }^{\dagger
}b_{i\sigma}=1$, as in the fermion or $d$-RVB theories \cite{ong_aspen}.
However, the presence of the nontrivial phase string operator, $\hat{\Theta}%
_{i\sigma}$, in (\ref{mutual}), makes this theory very different from the $d$%
-RVB theories. The phase string operator leads to nonlocal correlations
between the charge and spin degrees of freedom on account of hole motion in
a background with short range AF correlations. The fermion statistics of $%
c_{i\sigma }$ is also realized through the phase field 
\begin{equation}
e^{i\hat{\Theta}_{i\sigma }}=(\sigma )^{\hat{N}_h}(-\sigma )^i e^{i\Theta
_{i\sigma }^{\mathrm{string}}}~.
\end{equation}
The Klein factor $(\sigma )^{\hat{N}_h}$ guarantees anticommutation
relations between electron operators with different spins. Unless otherwise
stated, this factor will be omitted as it does not play any role in the
spin, transport, and single particle channels. The phase string operator $%
\Theta _{i\sigma }^{\mathrm{string}}$, is a nonlocal operator given by 
\begin{equation}
\Theta _{i\sigma }^{\mathrm{string}}\equiv \frac 12 \left[ \Phi _i^b-\sigma
\Phi _i^h\right]~,  \label{theta}
\end{equation}
where 
\begin{equation}
\Phi _i^b=\sum_{l\neq i}\theta _i(l) \left( \sum_\alpha \alpha n_{l\alpha
}^b -1\right)~,  \label{phib}
\end{equation}
and 
\begin{equation}
\Phi _i^h=\sum_{l\neq i}\theta _i(l)n_l^h~.  \label{phih}
\end{equation}
Here, $n_{l\alpha }^b$ and $n_l^h$ are spinon and holon number operators,
respectively. The quantity $\theta _i(l)$ is defined as $\theta _i(l)\equiv %
\mbox{Im ln $(z_i-z_l)$}$, where $z_l=x_l+iy_l$ is the complex coordinate of
a lattice site $l$.

Using the decomposition (\ref{mutual}), we find that the hopping term $H_t$
in (\ref{htj}) can be written as 
\begin{equation}
H_t=-t\sum_{\langle ij\rangle \sigma }\left( e^{iA_{ij}^s-i\phi
_{ij}^0}\right) h_i^{\dagger }h_j\left( e^{i\sigma A_{ji}^h}\right)
b_{j\sigma }^{\dagger }b_{i\sigma }+h.c.~,  \label{ht}
\end{equation}
where 
\begin{equation}
A_{ij}^s=\frac 12\sum_{l\neq i,j}\left[ \theta _i(l)-\theta _j(l)\right]
\left( \sum_\sigma \sigma n_{l\sigma }^b\right) ~,
\end{equation}
and 
\begin{equation}
\phi _{ij}^0=\frac 12\sum_{l\neq i,j}\left[ \theta _i(l)-\theta _j(l)\right]
~.
\end{equation}
The lattice gauge field $A_{ij}^s$ describes fictitious fluxoids bound to
spinons satisfying 
\begin{equation}
\sum_cA_{ij}^s=\pm \pi \sum_{l\in c}\left[ n_{l\uparrow }^b-n_{l\downarrow
}^b\right] ~,
\end{equation}
for an arbitrary closed path $c$ (assuming no spinons along $c$). The sign $%
\pm $ on the right hand side of the above equation refers to the clockwise
or counter clockwise choices for the path $c$. The quantity $\phi _{ij}^0$
describes a uniform flux threading through the 2D plane with a strength $\pi 
$ per plaquette, $\sum\nolimits_{\Box }\phi _{ij}^0=\pm \pi $. The gauge
field $A_{ij}^h$ is given by holons, 
\begin{equation}
A_{ij}^h=\frac 12\sum_{l\neq i,j}\left[ \theta _i(l)-\theta _j(l)\right]
n_l^h~,
\end{equation}
and describes fictitious fluxoids bound to holons, 
\begin{equation}
\sum_cA_{ij}^h=\pm \sum_{l\in c}n_l^h~.
\end{equation}
The superexchange term $H_J$ in (\ref{htj}) is given by 
\begin{equation}
H_J=-\frac J2\sum_{\langle ij\rangle \sigma \sigma ^{\prime }}\left(
e^{i\sigma A_{ij}^h}\right) b_{i\sigma }^{\dagger }b_{j-\sigma }^{\dagger
}\left( e^{i\sigma ^{\prime }A_{ji}^h}\right) b_{j-\sigma ^{\prime
}}b_{i\sigma ^{\prime }}~,  \label{hj}
\end{equation}
From equations (\ref{ht}) and (\ref{hj}), we see that the b-RVB theory can
be thought of as a $\pi $-flux theory where the $\pi $ fluxoids are bound to
the constituent particles, \emph{i.e.}, the spinons and the holons.

\subsection{Mean field solution}

The Hamiltonians for the charge and spin degrees of freedom, (\ref{ht}) and (%
\ref{hj}) respectively, can be solved within a generalized mean field
approximation. We expect the presence of the nonlocal fields $A^s_{ij}$ and $%
A^h_{ij}$ to cause residual effects of the short range AF correlations
within such a mean field treatment. We summarize below, a few results that
will be used in this paper. We refer the reader to an earlier paper by one
of us for details \cite{zy_99}.

A mean field decoupling leads to an effective Hamiltonian, $H_{\mathrm{eff}%
}=H_h+H_s$ , where the effective holon Hamiltonian 
\begin{equation}
H_h=-t_h\sum_{\langle ij\rangle }\left( e^{i[A_{ij}^s-\phi _{ij}^0]}\right)
h_i^{\dagger }h_j+h.c.~,  \label{hh}
\end{equation}
with $t_h\sim t$. The effective spinon Hamiltonian is given by 
\begin{equation}
H_s=-J_s\sum_{\langle ij\rangle \sigma }\left( e^{i\sigma A_{ij}^h}\right)
b_{i\sigma }^{\dagger }b_{j-\sigma }^{\dagger }+h.c.+\lambda \left(
\sum_{i\sigma }b_{i\sigma }^{\dagger }b_{i\sigma }-(1-\delta )N\right) ~,
\label{hs}
\end{equation}
where $J_s$ and $J$ are related by the generalized RVB order parameter $%
\Delta^s$ as $J_s =\frac{\Delta ^s}2J$. In the above expression, 
\begin{equation}
\Delta^s \equiv \Delta _{ij}^s=\sum_\sigma \left\langle e^{-i\sigma
A_{ij}^h}b_{i\sigma }b_{j-\sigma }\right\rangle ~.  \label{deltas}
\end{equation}
The single occupancy constraint is enforced on the average by the Lagrange
multiplier $\lambda $ in $H_s$. Note that the RVB order parameter $\Delta ^s$
above, incorporates the link field $A_{ij}^h$, and is invariant under the
following transformation: $A_{ij}^h\rightarrow A_{ij}^h+\theta _i-\theta _j$%
, and $b_{i\sigma }\rightarrow b_{i\sigma }e^{i\sigma \theta _i}$. Equations
(\ref{hh}) and (\ref{hs}) define the effective Hamiltonian, $H_{\mathrm{eff}%
} $. All the results presented in this paper are obtained using this
Hamiltonian.

The Hamiltonian, $H_J$ can be diagonalized by the Bogoliubov transformation, 
\begin{equation}
b_{i\sigma }=\sum_m\left( u_{m\sigma }(i)\gamma _{m\sigma }-v_{m\sigma}(i)
\gamma _{m-\sigma }^{\dagger } \right)~.  \label{bogo}
\end{equation}
The Bogoliubov factors, $u_{m\sigma}(i)$ and $v_{m\sigma}(i)$ can be
reexpressed in terms of a single particle wave function $w_{m\sigma}(i)$ as $%
u(v)_{m\sigma}(i) = u(v)_m w_{m\sigma}(i)$, and $w_{m\sigma}$ is determined
by the eigen equation 
\begin{equation}
\xi _mw_{m\sigma }(i)=-\frac J2\sum_{j=nn(i)} \Delta _{ij}^se^{i\sigma
A_{ij}^h}w_{m\sigma }(j)~.  \label{ew}
\end{equation}
%
The Hamiltonian, $H_s$ is then diagonal in the $\gamma_m$ representation, $%
H_s = \sum_{m\sigma} E_m \gamma^\dagger_{m\sigma} \gamma_{m\sigma}$, where $%
(E_m)_{min} \sim \delta J$, as shown in \cite{zy_99}. The RVB order
parameter is determined self consistently as 
\begin{equation}
\Delta _{ij}^s=\sum_{m\sigma } e^{-i\sigma A_{ij}^h}w_{m\sigma
}(i)w_{m\sigma }^{*}(j)(-u_mv_m) \left[ 1+\sum_\alpha \langle \gamma
_{m\alpha }^{\dagger }\gamma _{m\alpha }\rangle \right]~.  \label{deltasmf}
\end{equation}
In the following, we will always consider a real order parameter $\Delta
_{ij}^s$. In this case, it can be verified that $w_{m\sigma }=w_{m-\sigma
}^{*}$ and that $\left( \Delta _{ij}^s\right) ^{*}=\Delta _{ij}^s$. Finally,
it is useful to note that 
\begin{equation}
\langle \mathbf{S}_i\cdot\mathbf{S}_j\rangle =-\frac 1 2 |\Delta^s|^2<0
\end{equation}
for nearest neighbors, $ij$. This is consistent with the notion that the
b-RVB order parameter essentially characterizes the short range AF
correlations.

\section{Ground state: Superconducting state with $d$-wave symmetry}

The b-RVB ground state is underpinned by the RVB order parameter, $\Delta
_{ij}^s$. At low temperatures, the holons are condensed. At $T=0$, the
ground state is superconducting. In this section, we will show that the
superconducting order parameter has $d$-wave symmetry. This is due to the
short range AF correlations. When a hole moves from a lattice site $i+\hat{x}
$ to $i+\hat{y}$, the string operator (\ref{theta}) picks up a minus sign on
an average due to the short range AF correlations. We will show below that
this translates into $d$-wave pairing of electrons.

The pair operator between nearest neighbor sites $ij$, $\hat{\Delta} _{ij}^{%
\mathrm{SC}}\equiv \sum_\sigma \sigma c_{i\sigma }c_{j-\sigma }$, will now
be reexpressed using the boson representation (\ref{mutual}). For the
purposes of this section, we find it convenient to define the quantities, 
\[
\Phi _i^s=\sum_{l\neq i}\theta _i(l)\sum_\alpha \alpha n_{l\alpha }^b~, 
\]
and 
\[
\Phi _i^0=\sum_{l\neq i}\theta _i(l)~. 
\]
Then, it is easy to verify that 
\begin{equation}
\hat{\Delta}_{ij}^{\mathrm{SC}}= e^{i\frac 12\left( \Phi _i^s+\Phi
_j^s\right) }\hat{\Delta}_{ij}^0~,  \label{deltasc}
\end{equation}
where 
\begin{equation}
\hat{\Delta}_{ij}^0\equiv \left[ (-1)^j(-1)^{N_h}e^{-i\Phi _j^0-i\phi
_{ij}^0}\right] h_i^{\dagger }h_j^{\dagger }\hat{\Delta}_{ij}^s
\label{delta0}
\end{equation}

At low temperature, holons are Bose condensed, \emph{viz.}, $\left\langle
h_i^{\dagger }\right\rangle =h_0e^{-i\phi _i^h}$, where $\phi _i^h$ is a
fixed phase. Then, the superconducting order parameter can be written as a
product of an amplitude and a phase as 
\begin{equation}
\langle \hat{\Delta}_{ij}^{\mathrm{SC}}\rangle = \langle \hat{\Delta}%
_{ij}^0\rangle \langle e^{i\frac 12\left( \Phi _i^s+\Phi _j^s\right)
}\rangle ~.  \label{odlro}
\end{equation}
In the Appendix, we prove that the quantity $\langle \hat{\Delta}
_{ij}^0\rangle$ is a constant, independent of $ij$. The important physical
quantity in this discussion is the phase factor, $e^{i\frac 12\left( \Phi
_i^s+\Phi _j^s\right) }$ in (\ref{deltasc}).

First we note that a real superconducting order parameter will be
established only when phase coherence is realized; \emph{i.e.}, when spinons
are paired up in the ground state such that $\left\langle e^{i\frac 12\left(
\Phi _i^s+\Phi _j^s\right) }\right\rangle \neq 0$. Therefore, it should be
clear that the superconducting transition temperature $T_c$ is determined by
phase coherence \cite{vnm_02}.

Besides the phase (de)coherence, the topological phase factor $e^{i\frac 12
\left( \Phi _i^s+\Phi _j^s\right) }$in (\ref{deltasc}) also determines the $%
d $-wave symmetry of the order parameter. To see this, let us compare $e^{i
\frac 12\left( \Phi _i^s+\Phi _{i+\widehat{x} }^s\right) }$ and $e^{i\frac
12 \left( \Phi _i^s+\Phi _{i+\widehat{y} }^s\right) }$. 
%
First, we rewrite 
\[
e^{i\frac 12\left( \Phi _i^s+\Phi _{i+\widehat{\eta }}^s\right) }=e^{i\left(
A_{ii+\widehat{\eta }}^s+\Phi _{i+\widehat{\eta }}^s\right)}~, 
\]
which holds true when two spinons at $i$ and $i+\widehat{\eta }$ are
annihilated by $\hat{\Delta}_{ii+\widehat{\eta }}^0$ in (\ref{deltasc}). We
do so because $e^{iA_{ii+\widehat{\eta }}^s}$ is well defined with $%
\left\langle A_{ii+\widehat{\eta }}^s\right\rangle =0$ at $T=0$. Therefore, $%
\left\langle e^{iA_{ii+ \widehat{\eta }}^s}\right\rangle \approx e^{-\frac
12 \left\langle \left( A_{ii+ \widehat{\eta }}^s\right) ^2\right\rangle }>0$
in the ground state, and is independent of the nn link $ii+\hat{\eta}$. On
the other hand, the difference between $e^{i\Phi _{i+\widehat{x}}^s}$ and $%
e^{i\Phi _{i+\widehat{y}}^s}$ leads to a phase change. Consider 
\[
e^{i\Phi _{i+\widehat{x}}^s-i\Phi _{i+\widehat{y}}^s}=e^{i\left[ \theta _{i+ 
\widehat{x}}(i+\widehat{y})\sum_\sigma \sigma n_{i+\widehat{y}\sigma
}^b-\theta _{i+\widehat{y}}(i+\widehat{x})\sum_\sigma \sigma n_{i+\widehat{x}
\sigma }^b\right] +iA_{_{i+\widehat{x},i+\widehat{y}}}^s}~. 
\]
In the presence of short range AF correlations, one has $\sum_\sigma \sigma
n_{i+\widehat{y}\sigma }^b\approx \sum_\sigma \sigma n_{i+\widehat{x}\sigma
}^b$ such that 
\begin{eqnarray*}
e^{i\left[ \theta _{i+\widehat{x}}(i+\widehat{y})\sum_\sigma \sigma n_{i+ 
\widehat{y}\sigma }^b-\theta _{i+\widehat{y}}(i+\widehat{x})\sum_\sigma
\sigma n_{i+\widehat{x}\sigma }^b\right] } &\simeq &e^{i\left[ \theta _{i+ 
\widehat{x}}(i+\widehat{y})-\theta _{i+\widehat{y}}(i+\widehat{x})\right]
\sum_\sigma \sigma n_{i+\widehat{x}\sigma }^b} \\
&=&-1
\end{eqnarray*}
since $\theta _{i+\widehat{x}}(i+\widehat{y})-\theta _{i+\widehat{y}}(i+ 
\widehat{x})=\pm \pi$. We emphasize that such a such a sign change is
generally true on an average in an RVB background, and does \emph{not}
require long range AF order.

Thus, the presence of short range AF correlations among spins causes a $d$
-wave like sign change between $\left\langle \hat{\Delta}_{ii+\widehat{x} }^{%
\mathrm{SC}}\right\rangle$ and $\left\langle \hat{\Delta}_{ii+\widehat{y} }^{%
\mathrm{SC}}\right\rangle $. If one only retains the nn pairing amplitude $%
\left\langle \hat{\Delta}_{ii+\widehat{\eta }}^{\mathrm{SC}}\right\rangle
=\Delta ^0f_{i,i+\widehat{\eta }}$, with $f_{i,i+\widehat{x}}/f_{i,i+%
\widehat{y}}=-1$ , then in momentum space, one has the familiar $d$-wave
form factor 
\begin{equation}
\Delta _{\mathbf{k}}^{\mathrm{SC}}= 2\left\langle \hat{\Delta}_{ii+\widehat{x%
} }^{\mathrm{SC}}\right\rangle \left( \cos k_xa-\cos k_ya\right) \equiv
\Delta _{\mathbf{k} }^0  \label{D}
\end{equation}

\section{Elementary Excitations}

In this Section, we consider the confinement of bare spinon and holon
excitations. Though the superconducting state at $T=0$, $|\Psi _G\rangle $,
is characterized by spinon (RVB) pairing and a holon condensate, it does not
necessarily imply the presence of free spinon and holon excitations. In
fact, we will show that creating a bare spinon or holon excitation out of
the RVB vacuum costs infinite energy. On the other hand, we will show that
neutral $S=1$ excitations as well as quasiparticle excitations can be
created from the ground state with finite energies.

\subsection{Spin excitations}

A single spinon excitation can be constructed as $\gamma _{m\sigma
}^{\dagger }|\Psi _G\rangle $. As mentioned earlier, a mean field treatment
of (\ref{hj}) yields the result, $E_m\sim \delta J$. Therefore, we may be
tempted to conclude that the cost of creating a spinon excitation is finite.
However, the presence of the long range topological gauge field $A_{ij}^s$,
in the effective Hamiltonian for \emph{holons }(\ref{hs}) suggests that the
creation of a single spinon will be perceived as a half vortex, and
consequently would be a high energy excitation. To see this, consider the
energy cost of such an excitation, $\langle \Psi _G|\gamma _{m\sigma
}H_h\gamma _{m\sigma }^{\dagger }|\Psi _G\rangle -E_0^h$ , where $%
E_0^h=\langle \Psi _G|H_h|\Psi _G\rangle $. Now, consider $\langle \Psi
_G|\gamma _{m\sigma }e^{iA_{ij}^f}h_i^{\dagger }h_j\gamma _{m\sigma }|\Psi
_G\rangle $. From (\ref{bogo}), we see that 
\[
\gamma _{m\sigma }^{\dagger }=\sum_lw_{m\sigma }(l)\left[ u_mb_{l\sigma
}^{\dagger }+v_mb_{l-\sigma }\right] ~. 
\]
Using this, we find 
\[
\langle \Psi _G|\gamma _{m\sigma }e^{iA_{ij}^f}h_i^{\dagger }h_j\gamma
_{m\sigma }|\Psi _G\rangle =\sum_l\langle \Psi _G|e^{iA_{ij}^f}h_i^{\dagger
}h_j|\Psi _G\rangle e^{i\frac \sigma 2[\theta _i(l)-\theta
_j(l)]}|w_{m\sigma }(l)|^2~. 
\]
Noting that $\theta _i(l)-\theta _j(l)$ is independent of $l$ when summed
over all links $<ij>$ in $H_h$, and on using the normalization condition $%
\sum_l|w_{m\sigma }(l)|^2=1$, we get 
\[
\langle \Psi _G|\gamma _{m\sigma }H_h\gamma _{m\sigma }^{\dagger }|\Psi
_G\rangle -E_0^h=E_0^h\times \left[ \frac 1{2N}\sum_{<ij>}\cos \frac
12[\theta _i(l)-\theta _j(l)]-1\right] ~. 
\]
For a nearest neighbor link $<ij>$ far away from any lattice site $l$, one
has 
\[
|\theta _i(l)-\theta _j(l)|\sim \frac{a|\sin \alpha _{ij}|}{r_{li}} 
\]
where $\alpha _{ij}$ is the azimuthal angle of the link $<ij>$ and $%
r_{li}\sim r_{lj}$ is the distance between $l$ and $<ij>,$ and $a$ is the
lattice constant. Then, a summation over all nearest neighbor links $<ij>$
leads to 
\[
\frac 1{2N}\sum_{<ij>}\cos \frac 12[\theta _i(l)-\theta _j(l)]-1\sim \ln
\frac La~, 
\]
$L$ being the size of the sample. We thus see that it costs a
logarithmically divergent energy to create a single spinon excitation.
Consequently, single spinon excitations are not allowed in the
superconducting ground state at $T=0$. We reemphasize the crucial role
played by the topological gauge field $A_{ij}^s$, leading to this result.

The above analysis suggests that a logarithmic divergence in the excitation
energy can be avoided by pairing of spinons (binding of a vortex-antivortex
pair). Such an excitation is nothing but a neutral $S=1$ excitation. For
example, consider a spin excitation created by $S_i^z$, $S_i^z|\Psi
_G\rangle $. At $T=0$, the relevant term in $S_i^z|\Psi _G\rangle $ is $%
\gamma _{m\sigma }^{\dagger }\gamma _{m^{^{\prime }}-\sigma }^{\dagger
}|\Psi _G\rangle $. The excitation energy can be readily determined from the
mean field theory of $H_s$, as 
\[
\langle \Psi _G|\gamma _{m^{^{\prime }}-\sigma }\gamma _{m\sigma }H_s\gamma
_{m\sigma }^{\dagger }\gamma _{m^{^{\prime }}-\sigma }^{\dagger }|\Psi
_G\rangle -E_0^s=E_m+E_{m^{^{\prime }}}~, 
\]
which is finite. Now let us check the vortex energy in the charge sector, $%
H_h$. A straightforward calculation yields 
\begin{eqnarray*}
&&\langle \Psi _G|\gamma _{m^{^{\prime }}-\sigma }\gamma _{m\sigma
}H_h\gamma _{m\sigma }^{\dagger }\gamma _{m^{^{\prime }}-\sigma }^{\dagger
}|\Psi _G\rangle -E_0^h \\
&=&{\frac{E_0^h}{2N}}\sum_{<ij>}\sum_{l_1l_2}|w_{m\sigma
}(l_1)|^2|w_{m^{^{\prime }}-\sigma }(l_2)|^2\cos \left\{ \frac 12[\theta
_i(l_1)-\theta _j(l_1)]-\frac 12[\theta _i(l_2)-\theta _j(l_2)]\right\}
-E_0^h~.
\end{eqnarray*}
In obtaining the above, we have made repeated use of 
\[
h_ie^{\frac i2\Phi _j^h}=e^{\frac i2\theta _j(i)}e^{\frac i2\Phi _j^h}h_i~, 
\]
\[
b_{i\sigma }e^{\frac i2\Phi _j^b}=e^{i\frac \sigma 2\theta _j(i)}e^{i\frac
\sigma 2\Phi _j^b}b_{i\sigma }~, 
\]
\emph{etc.} Note that the factor 
\[
\frac 12[\theta _i(l_1)-\theta _j(l_1)]-\frac 12[\theta _i(l_2)-\theta
_j(l_2)] 
\]
vanishes when $l_1=l_2$. It has a dipole configuration when $l_1\neq l_2$, $%
l_1$ and $l_2$ being the cores of the spinon vortex-antivortex pair. The
corresponding energy is no longer logarithmically divergent, since the
factor $|w_{m\sigma }(l_1)|^2$ $|w_{m^{^{\prime }}-\sigma }(l_2)|^2$
restricts $l_1$ and $l_2$ within a characteristic length scale of the spinon
wave packet, $l_c$. Hence, we get (in the continuum limit) 
\begin{eqnarray*}
&&\langle \Psi _G|\gamma _{m^{^{\prime }}-\sigma }\gamma _{m\sigma
}H_h\gamma _{m\sigma }^{\dagger }\gamma _{m^{^{\prime }}-\sigma }^{\dagger
}|\Psi _G\rangle -E_0^h \\
&\sim &\frac{E_0^h}N\sum_{l_1l_2}|w_{m\sigma }(l_1)|^2|w_{m^{^{\prime
}}-\sigma }(l_2)|^2\ln \frac{|\mathbf{r}_{l_1}-\mathbf{r}_{l_2}|}{l_c}~.
\end{eqnarray*}
Thus, an $S=1$ spin excitation comprising a pair of spinons costs a finite
energy, and is an elementary excitation. It does not decay into free
spinons, due to a logarithmic confining force between the constituent
spinons. The excitation energy has a lower bound, $E_g\gtrsim 2E_s$, where $%
E_s=(E_m)_{\min }\sim \delta J$.

To conclude this subsection, we note that an $S=1$ excitation created by $%
S_i^{+}=(-1)^ib_{i\uparrow }^{\dagger }b_{i\downarrow }e^{i\Phi _i^h}$
should also have a finite excitation energy, as required by spin rotational
symmetry. This can be verified by an explicit calculation. In this case, the
relevant excitation at $T=0$ is of the form $\gamma _{m\uparrow }^{\dagger
}\gamma _{m^{^{\prime }}\uparrow }^{\dagger }e^{i\Phi _i^h}|\Psi _G\rangle $%
. Following the same steps as above, we find that the logarithmic divergence
is avoided by a cancellation between the $2\pi $ vortex $\Phi _i^h$ and two $%
\pi $ vortices introduced by $\gamma _{m\uparrow }^{\dagger }\gamma
_{m^{^{\prime }}\uparrow }^{\dagger }$. In the following subsection, we will
discuss an analogous effect which leads to a finite energy for the
quasiparticle excitation.

\subsection{Quasiparticle excitations}

A quasiparticle (or a quasihole) excitation is obtained as a linear
combination of $c_{l\sigma }|\Phi _G\rangle $ and $c_{l-\sigma }^{\dagger
}|\Phi _G\rangle $. We now show that the emergence and stability of such an
excitation is due to the confinement of bare spinon and holon excitations at 
$T=0$.

From (\ref{mutual}), we see that a quasihole excitation has the form 
\begin{equation}
c_{l\sigma }|\Psi _G\rangle \propto h_l^{\dagger }b_{l\sigma }e^{\frac
i2\left[ \Phi _l^s-\sigma \Phi _l^h\right] }|\Psi _G\rangle ~,  \label{qhole}
\end{equation}
\emph{viz.}, it comprises the holon, spinon and the vortex fields. Let us
examine these three components individually.

First, we have already seen that $b_{l\sigma }|\Psi _G\rangle $ would cost
an infinite energy in $H_h$. Similarly, a single holon added to the ground
state will also cost a logarithmically diverging energy in $H_s$. It is
easily seen that 
\begin{equation}
\frac{\langle \Psi _G|h_lH_sh_l^{\dagger }|\Psi _G\rangle }{\langle \Psi
_G|h_lh_l^{\dagger }|\Psi _G\rangle }-E_0^s=E_0^s\times \left[ \frac
1{2N}\sum_{<ij>}\cos \frac 12[\theta _i(l)-\theta _j(l)]-1\right] ~,
\label{s-holon}
\end{equation}
which diverges logarithmically with the lattice size. The third component in
(\ref{qhole}) is the vortex field. Consider 
\[
\langle \Psi _G|e^{+\frac i2\sigma \Phi _i^h}H_he^{-\frac i2\sigma \Phi
_i^h}|\Psi _G\rangle -E_0^h=E_0^h\times \left[ \frac 1{2N}\sum_{<ij>}\cos
\frac 12[\theta _i(l)-\theta _j(l)]-1\right] ~, 
\]
which is also logarithmically divergent. Similarly, 
\[
\langle \Psi _G|e^{-\frac i2\Phi _i^s}H_se^{\frac i2\Phi _i^s}|\Psi
_G\rangle -E_0^s=E_0^s\times \left[ \frac 1{2N}\sum_{<ij>}\cos \frac
12[\theta _i(l)-\theta _j(l)]-1\right] ~, 
\]
which diverges too. Therefore, in the ground state of the b-RVB, the
quasiparticle cannot decay into independent holon, spinon and vortex
excitations as it costs infinite energy. On the other hand, a quasiparticle
excitation as a composite excitation has a finite energy. In this case, a
cancellation of the vortex effects ensues on binding the holon and spinon
excitations to the vortex fields.

Consider $H_s$. Clearly, $b_{l\sigma }|\Phi _G\rangle$ costs a finite
energy, and we saw earlier that both $h_l^{\dagger }|\Phi _G\rangle $ and $%
e^{\frac i2 \Phi _l^s}|\Phi _G\rangle$ separately cost infinite energies in $%
H_s$. In contrast, consider the combination $h_l^{\dagger } e^{\frac i2\Phi
_l^s}$ in (\ref{qhole}). We will show below that the vortex effects cancel.
Now, 
\[
H_sh_l^{\dagger }e^{\frac i2\Phi _l^s}|\Psi _G\rangle = h_l^{\dagger }\left[
-J_s\sum_{\langle ij\rangle \sigma }\left( e^{i\sigma A_{ij}^h}\right)
b_{i\sigma }^{\dagger }b_{j-\sigma }^{\dagger }e^{i\sigma \chi _{ij}}+
h.c.\right] e^{\frac i2\Phi _l^s}|\Psi _G\rangle~, 
\]
where the phase twist $\chi _{ij}=1/2\left[ \theta _i(l)-\theta _j(l)\right]$
if ($ij$)$\neq l$ and $\chi _{ij}=0$ if $i=l$ or $j=l$, is responsible for
the logarithmic energy in (\ref{s-holon}). This is just the statement that
the holon is perceived as a half vortex through the term $A^h_{ij}$. Such a
vortex effect will be canceled by $e^{\frac i2\Phi _l^s}$, since 
\[
H_sh_l^{\dagger }e^{\frac i2\Phi _l^s}|\Psi _G\rangle =
h_l^{\dagger}e^{\frac i2\Phi _l^s} \left[ -J_s\sum_{\langle ij\rangle \sigma
}\left( e^{i\sigma A_{ij}^h}\right) b_{i\sigma }^{\dagger } b_{j-\sigma
}^{\dagger }+ h.c.\right] |\Psi _G\rangle~. 
\]
Equivalently, $\left[ H_s,\text{ }h_l^{\dagger }e^{\frac i2\Phi _l^s}\right]
=0 $, such that 
\[
\langle \Phi _G|h_le^{-\frac i2\Phi _l^s}H_sh_l^{\dagger }e^{\frac i2\Phi
_l^s}|\Phi _G\rangle =\langle \Phi _G|h_lh_l^{\dagger }H_s|\Phi _G\rangle 
\]
is finite. Consequently, $c_{i\sigma }|\Psi _G\rangle$ defined in (\ref
{qhole}) must have a finite matrix element in $H_s$.

A similar analysis can be carried out for the matrix elements of $H_h$. In
this case, $h_l^{\dagger }|\Psi _G\rangle$ and $e^{\frac i2\Phi _l^s}|\Psi
_G\rangle$ cost finite energies in $H_h$, but $b_{l\sigma }|\Psi _G\rangle$
and $e^{-\frac i2\sigma \Phi _l^h}|\Psi _G\rangle$ yield divergent energies.
The vortex effects can be cancelled by constructing the combination $%
b_{l\sigma}e^{-\frac i2\sigma \Phi_l^h}$, \emph{viz.}, $\left[ H_h,\text{ }%
b_{l\sigma }e^{-\frac i2\sigma \Phi _l^h}\right] =0~, $ which shows that the
vortex effect induced by $b_{l\sigma }$ in $H_h$ is cancelled by the effect
of $e^{-\frac i2\sigma \Phi _l^h}$.

Note that only the vortex phase factor $e^{\frac i2\left[ \Phi _l^s-\sigma
\Phi _i^h\right] }$ is single valued. Neither $e^{\frac i2\Phi _l^s}$ nor $%
e^{-\frac i2\sigma \Phi _l^s}$ are single valued by themselves \cite{zy_00}.
This means that $h_l^{\dagger }e^{\frac i2\Phi _l^s}$ and $b_{l\sigma
}e^{-\frac i2 \sigma \Phi _l^h}$ cannot be independent physical objects.
Therefore, the holon and spinon fields in (\ref{qhole}) have to be bound to
the vortex field $e^{\frac i2\left[ \Phi _l^s-\sigma \Phi _i^h\right] }$ as
a whole, such that the quasiparticle excitation is well defined and stable
in the superconducting state.

Now that we have demonstrated the stability of the quasiparticle excitation,
we determine its energy. Consider 
\begin{eqnarray*}
&&\left[ H_s,\text{ }c_{i\sigma }\right] \\
&=&-J_s\sum_{m=nn(i)}\left[ \sum_\alpha b_{i\alpha }^{\dagger }b_{m-\alpha
}^{\dagger }e^{i\alpha A_{im}^h}+h.c.\right] c_{i\sigma } \\
&&+c_{i\sigma }J_s\sum_{m=nn(i)}\left( \sum_\alpha b_{i\alpha }^{\dagger
}b_{m-\alpha }^{\dagger }e^{i\alpha A_{im}^h}+h.c.\right) \\
&=&-J_s\sum_{m=nn(i)}h_i^{\dagger }e^{i\hat{\Theta}_{i\sigma }}\left(
\sum_\alpha b_{i\alpha }^{\dagger }b_{m-\alpha }^{\dagger }e^{i\alpha
A_{im}^h}e^{i\frac \alpha 2\theta _i(m)}+h.c.\right) b_{i\sigma } \\
&&+J_s\sum_{m=nn(i)}h_i^{\dagger }e^{i\hat{\Theta}_{i\sigma }}b_{i\sigma
}\left( \sum_\alpha b_{i\alpha }^{\dagger }b_{m-\alpha }^{\dagger
}e^{i\alpha A_{im}^h}+h.c.\right)
\end{eqnarray*}
Each term in the above expression involves three spinon operators and can be
linearized using the mean field approximation (\ref{deltas}). Then, 
\[
\left[ H_s,\text{ }c_{i\sigma }\right] \simeq 10J_s\Delta ^sc_{i\sigma }%
\text{ \ \ }+J_s\sum_{m=nn(i)}h_i^{\dagger }e^{i\hat{\Theta}_{i\sigma
}}b_{m-\sigma }^{\dagger }e^{i\sigma A_{im}^h}\left\langle b_{i\sigma
}b_{i\sigma }^{\dagger }\right\rangle 
\]
(noting that terms involving $e^{i\frac \sigma 2\theta _i(m)}$ vanish after
averaging over all possible $\theta _i(m))$. The second term on the rhs of
the above equation can be rewritten, using (\ref{mutual}) and (\ref{deltasc}%
). We get 
\[
\left[ H_s,\text{ }c_{i\sigma }\right] \approx 10\Delta ^sJ_sc_{i\sigma }-%
\frac{J_s}{\Delta ^s}\sum_{m=nn(i)}\left\langle b_{i\sigma }b_{i\sigma
}^{\dagger }\right\rangle \left( \frac{\Delta _{im}^{\mathrm{SC}}}{h_0^2}%
\right) \sigma c_{m-\sigma }^{\dagger }~. 
\]
In obtaining the above expression, we have approximated $h_i^{\dagger }$ by $%
h_m$, since the holons are condensed with an expectation value, $<h^{\dagger
}>=h_0$ (its phase $\phi ^h$ can be always set as zero due to the gauge
choice discussed in Appendix). The factor of $\sigma $ in the second term of
the rhs comes from commuting the holon field $h_m$ with $e^{-i\hat{\Theta}%
_{m-\sigma }}$. We have used the definition of the superconducting order
parameter, 
\begin{eqnarray*}
\Delta _{im}^{\mathrm{SC}} &=&\sum_\sigma \left\langle h_i^{\dagger
}h_j^{\dagger }e^{-i\sigma A_{ij}^h}b_{i\sigma }b_{j-\sigma }(-\sigma
)^i(\sigma )^m(-1)^{N_h}e^{\frac i2\left( \Phi _i^b+\Phi _m^b\right)
}\right\rangle ~ \\
&\simeq &h_0^2\Delta ^s\left\langle (-\sigma )^i(\sigma
)^m(-1)^{N_h}e^{\frac i2\left( \Phi _i^b+\Phi _m^b\right) }\right\rangle ~.
\end{eqnarray*}

Then in momentum space, 
\begin{equation}
\left[ H_s,\text{ }c_{\mathbf{k}\sigma }\right] \approx 10\Delta ^sJ_sc_{%
\mathbf{k}\sigma }-\Delta _{\mathbf{k}}^d\sigma c_{-\mathbf{k}-\sigma
}^{\dagger }~,  \label{bdgs}
\end{equation}
where $\Delta _{\mathbf{k}}^d=\frac{(3-\delta )J}4\left( \frac{\Delta _{%
\mathbf{k}}^0}{h_0^2}\right) $. Here, we have used $\left\langle b_{i\sigma
}b_{i\sigma }^{\dagger }\right\rangle =1+\left\langle b_{i\sigma }^{\dagger
}b_{i\sigma }\right\rangle =(3-\delta )/2$ and $J_s=\frac{\Delta ^s}2J$.

On the other hand, the hopping term will generally give rise to a kinetic
energy\cite{remark00}

\begin{equation}
\left[ H_h,\text{ }c_{\mathbf{k}\sigma }\right] =-(\epsilon _{\mathbf{k}%
}-\mu )c_{\mathbf{k}\sigma }  \label{bdgh}
\end{equation}
with $\epsilon _{\mathbf{k}}\sim t_h (\cos k_x+\cos k_y)$. Clearly, a
quasiparticle operator can be constructed in the usual manner as 
\[
\alpha _{\mathbf{k}\sigma }^{\dagger }=\widetilde{u}_{\mathbf{k}}c_{\mathbf{k%
}\sigma }^{\dagger }-\sigma \widetilde{v}_{\mathbf{k}}c_{-\mathbf{k}-\sigma
} 
\]
where $\widetilde{u}_{\mathbf{k}}^2+\widetilde{v}_{\mathbf{k}}^2=1$. On
imposing the condition $[H_s,\text{ }\alpha _{\mathbf{k}\sigma }^{\dagger
}]=E_{\mathbf{k}}^{qp}\alpha _{\mathbf{k}\sigma }^{\dagger }$ and using
equations (\ref{bdgs}) and (\ref{bdgh}), it is straightforward to get the
standard expressions, $E_{\mathbf{k}}^{qp}=\sqrt{(\epsilon _{\mathbf{k}}-\mu
)^2+(\Delta _{\mathbf{k}}^d)^2}$, $\widetilde{u}_{\mathbf{k}}=\frac 1{\sqrt{2%
}}\left( 1+\frac{\epsilon _{\mathbf{k}}-\mu }{E_{\mathbf{k}}^{qp}}\right)
^{1/2}$, and $\widetilde{v}_{\mathbf{k}}=\frac 1{\sqrt{2}}\left( 1-\frac{%
\epsilon _{\mathbf{k}}-\mu }{E_{\mathbf{k}}^{qp}}\right) ^{1/2}\mathrm{sgn}%
(\Delta _{\mathbf{k}}^d)$. The first term on the rhs of (\ref{bdgs}) can be
absorbed in the definition of the chemical potential, $\mu $. The latter can
be determined within mean field theory in the usual manner, by fixing the
particle number. 

The d-wave gap $\Delta _{\mathbf{k}}^d$ vanishes along the two nodal lines, $%
k_x=\pm k_y$. Thus, the quasiparticle spectrum has four nodes at $\epsilon
_{k_f}-\mu =0$.

It is important to note that even though the confining force between the
holon, spinon, and vortex constituents within a quasiparticle is weak
(logarithmic), the quasiparticle should be a well defined, long lived
excitation. Even if we neglect the logarithmic energy cost that keeps them
bound (which becomes important when the components are separated far away),
the mean field energies for holon and spinon constituents are zero and $E_m$%
, respectively. Thus, for $E_{\mathbf{k}}^{qp}<\left( E_m\right) _{\min
}\equiv E_s$, the quasiparticle energy is always lower than its
constituents. This implies that the nodal quasiparticle excitation will be
observed as a sharp spectral feature. Note that the vortex field in (\ref
{qhole}) also provides a confining potential that glues the spin and charge
degrees of freedom. This lowers the quasiparticle energy, in comparison to
the mean field energies of the spinon and holon excitations.

However, at higher energies, the signature of the composite quasiparticle
will be seen in the spectral function. For energies $\omega >E_s$, the
spectral function, $A(\mathbf{k},\omega )=-\frac 1\pi \mathrm{Im} G(\mathbf{k%
},\omega )\mathrm{sgn}(\omega )$, displays a composite structure in the form
of a broad peak and spectral weight at higher energies \cite
{vnm_02arpes,zy_00}. Therefore, even though low energy quasiparticles in the
b-RVB superconductor seemingly look like their counterparts in a
conventional d-wave BCS superconductor, they are easily distinguished from
the latter at higher energies, by a broad incoherent structure. This
reflects the composite nature of the RVB vacuum (\emph{i.e.}, spin charge
separation).

It is clear that the width of the quasiparticle would increase if $E_{%
\mathbf{k}}^{qp}$ becomes comparable to, or greater than, $E_s$. Typically,
this would first happen around the corners of the Brillouin zone, ($\pm \pi
,0)$ and $(0,\pm \pi )$. In these regions, $\Delta _{\mathbf{k}}^d$ and $E_{%
\mathbf{k}}^{qp}$ attain their maximal values. Since the mean field
excitation energies of the holon and spinon constituents are now comparable
to, or smaller than, $E_{\mathbf{k}}^{qp}$, the quasiparticle will be seen
as a loosely bound spinon-holon composite. This is consistent with results
from exact diagonalization \cite{martins_99,tohyama_00}.

A similar effect also occurs as the hole concentration is reduced. As $%
\delta \rightarrow 0$, $\Delta _{\mathbf{k}}^d$ is finite (recall the factor
of $h_0^2$ in the denominator of $\Delta _{\mathbf{k}}^d$), whereas $E_s\sim
J\delta \rightarrow 0$. Therefore, at small $\delta $, the sharp
quasiparticle peak evolves into a broad feature quickly as one moves away
from the nodal regions to the corners of the Brillouin zone. Eventually, for
low enough doping, the spectral function becomes very broad away from the
nodal points $\left( \pm \pi /2,\pm \pi /2\right) $, and is best described
in terms of spin charge separation rather than a bound excitation.
Elsewhere, we discussed this in detail for the case of a single hole doped
into the antiferromagnetic Mott insulator \cite{zy_01}. 

\section{Summary}

In this paper, we discussed the ground state and the nature of excitations
in a bosonic resonating valence bond (b-RVB) superconductor. We considered
an effective Hamiltonian for the spin and charge degrees of freedom. The
effective Hamiltonian was derived by rewriting the $t-J$ model using the
phase string decomposition. This decomposition can be thought of as a
bosonization scheme for the $t-J$ model. The ground state at $T=0$ is a
d-wave superconducting state which arises from RVB pairing and holon
condensation.

The physical reason behind $d$-wave pairing has a simple explanation.
Suppose one of the holes forming the Cooper pair is located at site $i$. As
the other hole belonging to the pair is transported from site $i+\hat{x}$ to 
$i+\hat{y}$, say, \emph{via} $i+\hat{x}+\hat{y}$ (the shortest path), it
picks pick up a phase string, $\sigma _{i+\hat{x}+\hat{y}}\sigma _{i+\hat{y}%
} $, due to an irreparable displacement of the Marshall sign that occurs due
to hole motion \cite{zy_00}. Owing to the short range AF correlations, $%
\sigma _{i+\hat{x}+\hat{y}}\sigma _{i+\hat{y}}<0$ on the average, which
implies that the Cooper pair at sites $i$ and $i+\widehat{y}$ changes sign
relative to the pair at $i$ and $i+\hat{x}$.

The superconducting condensation in the b-RVB and conventional d-RVB
theories \cite{ong_aspen} both involve the RVB pairing of spinons and holon
condensation. The main distinction between these two classes of theories
lies in the origins of kinetic energy frustration. In the b-RVB theory, the
kinetic energy of holons is strongly frustrated, due to the short range AF
correlations that are carefully incorporated. This kinetic energy
frustration is realized as the phase string effect. A partial recovery of
the frustrated kinetic energy occurs at the superconducting transition.
However, spinon and holon excitations out of the b-RVB vacuum still exhibit
the signature of strong mutual frustration between themselves: a single
spinon or holon excitation always invites a severe response from the other
degrees of freedom, causing a logarithmically divergent energy. This is the
origin of the low energy quasiparticle, by spin charge recombination.

We showed that in the superconducting state, $S=1$ excitations composed of
spinon pairs have finite energies, although single spinons would cost a
logarithmic energy. Similarly, a composite quasiparticle excitation,
carrying both spin and charge, is well defined and has a finite energy.
Though the quasiparticle operator we constructed has a superficial
similarity with the nodal quasiparticle in a conventional $d$ -wave BCS
superconductor, there are important differences. The first is that neutral $%
S=1$ excitations in the b-RVB state, to leading order, are independent of
quasiparticle excitations. They occur as bound spinon pairs, whereas in BCS
theory, the $S=1$ excitation is built out of quasiparticles. A second
difference is that the composite character of the quasiparticle can be seen
in the spectral function, at high energies. For low doping, even the low
energy quasiparticle is unstable at the Brillouin zone corners. It should be
noted that the quasiparticle in the b-RVB theory becomes incoherent at $T_c$%
. As discussed in Ref. \cite{vnm_02}, the superconducting transition occurs
when spinons are deconfined such that free spinon-vortices destroy the phase
coherence of the order parameter (\ref{odlro}). The proliferation of spinon
vortices leads to the screening of the logarithmic confining potential, and
the quasiparticle decays into its constituents. These differences in the
excitation spectra reflect the fundamental difference between a BCS
superconductor and an RVB superconductor. The former is constructed from a
Fermi sea of electrons, whereas the latter is built from an RVB vacuum which
describes a doped antiferromagnetic Mott insulator.

Though spin charge separation is the key feature in the description of the
doped Mott insulator, an important point we discussed in this paper is that
bare holons and spinons need not necessarily exist as excitations of the RVB
vacuum. The fact that a single holon or spinon cannot be created or
annihilated in the system at $T=0$
is consistent with the requirement of the no
double occupancy constraint at the global level. In contrast, an $S=1$
spin excitation or a quasiparticle excitation does not violate
such a constraint. The confinement of single spinon and holon excitations,
along with the emergence of $S=1$ spin excitations and quasiparticles as the
low lying elementary excitations in the b-RVB theory has furthered our
understanding of the concepts of the RVB state and spin charge separation.
In fact, deconfinement of spinons and holons plays a key role in determining
the superconducting transition and in the description of the high
temperature phases of the b-RVB theory. The phase string decomposition (\ref
{mutual}) provides a natural realization of a two dimensional fermion-boson
transmutation envisaged by Baskaran, during the early years of the RVB
theory \cite{baskaran_88}. Recently, Sarker has discussed the recombination
of spin and charge degrees of freedom in the superconducting state of the $%
t-J$ model, using a Schwinger boson representation \cite{sarker_00}.
Spin-charge recombination in the superconducting phase was also conjectured
previously by Wen, Lee and others \cite{wen_96} within the slave-boson
framework. In contrast, fractionalization of electrons or spin charge
separation of the elementary excitations persists in the superconducting
state obtained within Z$_2$ gauge theory \cite{senthil_00}.

Finally, in this paper, we were mainly interested in the stability and
finite energy of the nodal quasiparticle. We showed that the composite
character of this excitation can be observed clearly as a function of
doping. As the doping is reduced, the quasiparticle peak around the corners
of the Brillouin zone would broaden and the spectral function is better
described in terms of spin charge separation. A quantitative analysis of
this phenomenon in terms of the life time of the quasiparticle excitation,
and a comparison with experimental results available on the evolution of the
Fermi surface with doping, will be discussed elsewhere.

\acknowledgments
We thank P. W. Anderson and N. P. Ong for urging us to clarify the nature of
the fermionic quasiparticle excitation in the b-RVB theory. Y.Z. and Z.Y.W.
acknowledge partial support from NSFC Grant 90103021.

\appendix{}

\section{Appendix: Evaluation of $\left\langle \hat{\Delta}%
_{ij}^0\right\rangle $}

The amplitude of the superconducting order parameter in (\ref{odlro}) can be
expressed by

\begin{equation}
\left\langle \hat{\Delta}_{ij}^0\right\rangle =g_{ij}h_0^2\Delta ^s
\end{equation}
with

\begin{equation}
g_{ij}\equiv (-1)^j(-1)^{N_h}e^{-i(\Phi _j^0+2\phi _j^h)-i\left[ \phi
_{ij}^0+(\phi _i^h-\phi _j^h)\right] }  \label{f}
\end{equation}

The link quantity $g_{ij}$ defined in (\ref{f}) may be rewritten as:

\begin{equation}
g_{ij}=(-1)^j(-1)^{N_h}e^{-i(\Phi _j^0+2\phi _j^h)}e^{-i\overline{\phi }%
_{ij}^0}
\end{equation}
with $\overline{\phi }_{ij}^0\equiv \phi _{ij}^0+(\phi _i^h-\phi _j^h).$

By using $\theta _i(j)-\theta _j(i)=\pm \pi ,$ one can generally show that

\[
g_{ij}=g_0e^{-i\sum_c2\overline{\phi }_{lm}^0}e^{-i\overline{\phi }_{ij}^0} 
\]
where $g_0\equiv (-1)^{j_0}(-1)^{N_h}e^{-i(\Phi _{j_0}^0+2\phi _{j_0}^h)}$
and $c$ represents an arbitrary path composed of nn links on the lattice
which connects the site $j$ and a reference site $j_0$. Note that, as $%
\sum_{\Box }2\overline{\phi }_{ij}^0=\sum_{\Box }2\phi _{ij}^0=\pm 2\pi $, $%
g_{ij}$ is independent of the choice of the path $c$.

Once $j_0$ is fixed, $g_{ij}$ on every link will be determined. For example,
on the horizontal bonds connected to $j_0$: ($j_0$, $j_0\pm \widehat{x}$), ($%
j_0$, $j_0\pm 2\widehat{x}$), ($j_0$, $j_0\pm 3\widehat{x}$), ...

\begin{eqnarray}
g_{j_0,j_0\pm \widehat{x}} &=&g_0e^{-i2\overline{\phi }_{j_0\pm \widehat{x}%
,j_0}^0-i\overline{\phi }_{j_0,j_0\pm \widehat{x}}^0}=g_0e^{i\overline{\phi }%
_{j_0,j_0\pm \widehat{x}}^0}  \nonumber \\
g_{j_0\pm \widehat{x},j_0\pm 2\widehat{x}} &=&g_0e^{-i(2\overline{\phi }%
_{j_0\pm 2\widehat{x},j_0\pm \widehat{x}}^0+2\overline{\phi }_{j_0\pm 
\widehat{x},j_0}^0)-i\overline{\phi }_{j_0\pm \widehat{x},j_0\pm 2\widehat{x}%
}^0}  \nonumber \\
&&......
\end{eqnarray}
Similarly on the vertical bonds connected to $j_0$: ($j_0$, $j_0\pm \widehat{%
y}$), ($j_0$, $j_0\pm 2\widehat{y}$), ($j_0$, $j_0\pm 3\widehat{y}$), ..., 
\begin{eqnarray}
g_{j_0,j_0\pm \widehat{y}} &=&g_0e^{-i2\overline{\phi }_{j_0\pm \widehat{y}%
,j_0}^0-i\overline{\phi }_{j_0,j_0\pm \widehat{y}}^0}=g_0e^{i\overline{\phi }%
_{j_0,j_0\pm \widehat{y}}^0}  \nonumber \\
g_{j_0\pm \widehat{y},j_0\pm 2\widehat{y}} &=&g_0e^{-i(2\overline{\phi }%
_{j_0\pm 2\widehat{y},j_0\pm \widehat{y}}^0+2\overline{\phi }_{j_0\pm 
\widehat{y},j_0}^0)-i\overline{\phi }_{j_0\pm \widehat{y},j_0\pm 2\widehat{y}%
}^0}  \nonumber \\
&&......
\end{eqnarray}

Now let us consider the choice of the gauge in $\overline{\phi }_{ij}^0$
which satisfies 
\begin{equation}
\sum_{\Box }\overline{\phi }_{ij}^0=\pm \pi  \label{pi}
\end{equation}
The holon Hamiltonian in the holon condensed phase reduces to 
\begin{equation}
H_h=-t_hh_0^2\sum_{\langle ij\rangle }e^{i\left[ A_{ij}^s-\overline{\phi }%
_{ij}^0\right] }+h.c.  \label{hh0}
\end{equation}
Note that in the ground state all spinons are paired up such that $A_{ij}^s$
is cancelled out and $\left\langle e^{iA_{ij}^s}\right\rangle \approx
e^{-\frac 12\left\langle \left( A_{ij}^s\right) ^2\right\rangle }$ which is
generally independent of ($ij).$ Then 
\begin{equation}
\left\langle H_h\right\rangle =-t_hh_0^2\left\langle
e^{iA_{ij}^s}\right\rangle \sum_{\langle ij\rangle }e^{-i\overline{\phi }%
_{ij}^0}+h.c.
\end{equation}
which is optimized at the staggered-$\pi $-flux gauge choice of

\begin{eqnarray}
\overline{\phi }_{ii+\widehat{x}}^0 &=&\pm \left( -1\right) ^i\pi /4 
\nonumber \\
\overline{\phi }_{ii+\widehat{y}}^0 &=&\mp \left( -1\right) ^i\pi /4
\label{staggered}
\end{eqnarray}
satisfying (\ref{pi}). In (\ref{staggered}), there are two choices of signs, 
$\pm ,$ for the staggered $\pi $ flux at a given plaquette.

On averaging over these two equivalent gauge choices, we get 
\begin{eqnarray*}
g_{j_0,j_0\pm \widehat{x}} &=&g_0\left\langle e^{i\overline{\phi }%
_{j_0,j_0\pm \widehat{x}}^0}\right\rangle =g_0/\sqrt{2} \\
g_{j_0\pm \widehat{x},j_0\pm 2\widehat{x}} &=&g_0\left\langle e^{-i\overline{%
\phi }_{j_0\pm \widehat{x},j_0\pm 2\widehat{x}}^0}\right\rangle =g_0/\sqrt{2}
\\
&&...
\end{eqnarray*}
In general

\begin{equation}
g_{ij}=g_0/\sqrt{2}
\end{equation}


\begin{thebibliography}{99}
\bibitem{remark00}  The evaluation of the commutator for the kinetic term is
similar to the calculation done in Ref. \cite{zy_00}, in which a phase shift
in the single-particle wave function $w_{m\sigma }(i)$ must be used when a
holon changes positions. In Ref. \cite{zy_00}, the original hopping term of
the $t-J$ Hamiltonian instead of $H_h$ was used.
\end{thebibliography}
\end{document}